# Building Confidence not to be Phished through a Gamified Approach: Conceptualising User's Self-Efficacy in Phishing Threat Avoidance Behaviour

Gitanjali Baral and Nalin Asanka Gamagedara Arachchilage
School of Engineering and Information technology
University of New South Wales at Canberra, Australia
g.baral@student.adfa.edu.au
nalin.asanka@adfa.edu.au

**Abstract.** Phishing attacks are prevalent and humans are central to this online identity theft attack, which aims to steal victims' sensitive and personal information such as username, password, and online banking details. There are many anti-phishing tools developed to thwart against phishing attacks. Since humans are the weakest link in phishing, it is important to educate them to detect and avoid phishing at- tacks. One can argue self-efficacy is one of the most important determinants of individual's motivation in phishing threat avoidance behavior, which has co-relation with knowledge. The proposed research endeavors on the user's self-efficacy in order to enhance the individual's phishing threat avoidance behavior through their motivation. Using social cognitive theory, we explored that various knowledge attributes such as observational (vicarious) knowledge, heuristic knowledge and structural knowledge contributes immensely towards the individual's self-efficacy to enhance phishing threat prevention behavior. A theoretical framework is then developed depicting the mechanism that links knowledge attributes, self-efficacy, threat avoidance motivation that leads to users' threat avoidance behavior. Finally, a gaming prototype is designed in-cooperating the knowledge elements identified in this research that aimed to enhance individual's self-efficacy in phishing threat avoidance behavior.

## 1 Introduction

The "Wannacry Ransomeware" attack, in year 2017, is one of the biggest cyber attack in 20th century which has affected 150 countries worldwide. This crypto worm infected more than 230,000 computers with over 200,000 victims held for ransom of $300-$600 (via bitcoin) [19]. This Wannacry Ransomware initially started propagating through a phishing attack (i.e. phishing email). Phishing is an online identity theft, in which, the perpetrator attempts to steal victims' sensitive and personal information such as username, password, online banking details, personal and professional credentials [11] [13] [18]. Phishing attacks are carried out through various email tricking (a.k.a email spoofing) and website mimicking (a.k.a website spoofing) methods for social, financial and organisational secret service frauds [24, 30]. As the look and feel of both legitimate and fake websites and emails are same, it is quite challenging to differentiate between them, thereby making users prone to this type of attack. With the advancement in technology and widespread use of internet enabled devices, new phishing techniques are introduced every day and this trend may continue for many years to come [34, 37].

Recently, many anti-phishing solutions have been proposed and developed to avoid and regulate phishing attack and also to provide protection to computer users [10]. These phishing detection approaches can be broadly categorized into three major categories: Software-based methods; User warning methods; and Knowledge-based methods [28]. However, the Software-based, User warning, and Knowledge-based approaches using computing systems are not self-operated and thus must be functioned by its users. In addition, it is the user that provides inputs to the computer systems and accesses its expected output by making themselves as its integral part of this computing system as well as phishing solution. It is, therefore, necessary to educate users how to differentiate legitimate and illegitimate websites. Web-based training, interactive game-based training, contextual training or spear phishing experiments, embedded training (i.e. through machine and software that people normally use) and non-embedded training are some of the anti-phishing awareness methods used by the researchers to build and enhance user's knowledge about cyber threats [5].

Classroom based discussions and training [33] and web-based ant-phishing training materials from organisations such as eBay, Microsoft, etc [38] are some of the methods used to impart anti-phishing training to the user. Recently, Hanna Alghamdi [6] developed a phishing quiz mobile application to check the phishing detection ability of the user and to assess efficacy of online anti-phishing educational materials. However, the study results acknowledged that the users found trouble in understanding the phishing avoidance educational materials. The author argued that some factors like language of the learning material, complexity and difficult terminology used in the teaching material could be possible contributing factor. The author also argued that it is essential to simulate real life phishing attacks via the phishing quiz applications to observe peoples' reactions in the environment. Study conducted by authors in [7] have found that teaching phishing education does not always make the user better at recognising what is a phish and differentiating between a legitimate and fake websites due to their increased level of suspiciousness. Kumaraguru et al. [28] designed and evaluated the two educational interventions for their embedded training system; one containing warning with actionable items such as text and graphics, and the other containing comic strip format to deliver similar data. Their result demonstrated that embedded training was more effective with the comic characters than the previous format of security notices. Authors also noted that further implemen-tation of the embedded system with different design features such as short video of a story is also possible.

Game-based learning is another highly interactive approach for educating computer users to gain knowledge of phishing attacks while playing the game [8] [13]. Authors in [20] designed a cyber video game, CyberCIEGE, to provide security training to employees in an interactive and engaging format that acted as a security awareness tool. The result of their study clearly demonstrated the effectiveness as well as flexibility of video games in both general and organisational context. Authors in [9] designed an anti-phishing mobile gaming tool for computer users to protect themselves from phishing attacks. The game was aimed to enhance the user's avoidance behaviour through their motivation to secure themselves from phishing attacks. The authors demonstrated the efficacy of game based tools in enhancing the motivation of users in avoiding phishing attacks. Authors in [36] conducted a comparative study to determine the efficacy of different anti-phishing training such as playing a video game, reading the tutorial of the game and reading online training materials. Authors found that the participants who played the video game were better prepared in thwarting the phishing attack. Authors in [14] designed a prototype of a mobile game to enhance user's threat avoidance behaviour through their motivation to combat against phishing attacks. The study was conducted among 40 participants with pre and post-test, where 20 test takers played the mobile game and the rest 20 were asked to read the website directly. The research results illustrated that mobile game participants were more successful in identifying the illegitimate website than the participants who read the website directly. The researchers in this study believed that the concept of teaching users through this mobile game helps them to avoid and protect themselves from phishing attacks. The advantages of this game-based learning methods are that it is more challenging, motivating and engaging and thus, it delivers more knowledge with a better outcome [3].

We, the end users or the humans, are the weakest link in the security chain [27]. Due to this human factor, vulnerability exists in the system which is targeted by the attacker. Phishing attackers take advantage of this human vulnerability (i.e., mistake made by the end user while operating a computing system) rather than system vulnerabilities (i.e., flaw or weakness present in a system) by observing the human interaction with the computers or by portraying messages [35]. This human vulnerability may occur due to the lack of confidence and knowledge while using a computing system and anti-phishing tool [12]. For example, it is always difficult to identify the potential phishing attacks without the confidence and knowledge of using the computing system. Unfortunately, computing systems are not self-operated and thus must be functioned by its users. It is, therefore, necessary to be confident while the end-user interacts with the computer system and making security trust decisions. It is also true that these machines are unlikely to perform flawlessly without proper user interfaces, user knowledge and confidence. Previous research has revealed that there is a co-relation between user's knowledge and their self-efficacy. Users with more knowledge of phishing attacks are the ones who are more confident in thwarting phishing attacks [12].

This research aims to incorporate the user's confidence into a Game based learning tool, which will create and enhance user awareness through their phishing threat avoidance behaviour through their motivation. The proposed research will help users to build their confidence in a knowledge-based way so that the occurrence of such phishing attacks can be minimised or prevented. Thus, this research intends to determine if knowledge attributes such as observational knowledge and past-experience knowledge influence on computer users self-efficacy to combat against phishing attacks. The contributions of this research are; development of a theoretical model to correlate different knowledge attributes to users' self-efficacy; and development of gaming scenarios to conceptualise this theoretical model to a game based learning tool. This research focuses on investigating the knowledge factors or attributes that enhance the user's self-efficacy in phishing threat avoidance behaviour. The proposed research would lead to enhancement of user confidence through their increased awareness and phishing avoidance behaviour. The outcome of the proposed research would contribute in the reduction in number of phishing attacks. The rest of this paper is organised as follows. A background literature survey on the proposed research is provided in Section 2; theoretical framework is developed and presented in Section 3; while, Section 4 presents a game design with a story that explains how one can integrate the elements derived from our proposed theoretical model. The proposed game design teaches people how to thwart phishing attacks. The conclusion and future work is provided in Section 5.

## 2    Background Literature

Consideration of users' self-efficacy factor while developing the user educational anti-phishing tool is very nominal in the existing literature [6, 20, 28, 33]. In addition, existing literature from previous research has also revealed that there is a strong relationship between people's self-efficacy and their active involvement and achievements [18]. Bandura in [17] defined self-efficacy as one's own judgments about their capabilities to establish and execute actions to acquire certain target. Bandura also argued that people are more active to put themselves in certain behaviours if they think they are confident in executing those behaviours successfully. According to authors in [10], it is necessary to build threat perception in user so that they will motivate themselves to combat phishing attack through their avoidance behaviour. Self-efficacy leads to individual's better threat perception and such threats are available on the cyber space. If you are more knowledgeable about the phishing threats, your threat perception would be significantly high. As identified by authors in [35], cyber criminals target and exploit human vulnerability rather than the system vulnerability. Therefore, user-centred security educational tools should consider the user self-efficacy factor that will directly motivate them to perceive threat while working on cyber-space to avoid phishing attacks. Being knowledgeable gives rise to our inner confidence. Hence, the more knowledge we have, the more confident we are.

Authors in [29] proposed a Technology Threat Avoidance Theory (TTAT) model to describe how individual IT user's are motivated to avoid an IT threat by taking a safeguarding measure when they perceive an IT threat exists. The users' believe that the threat can be avoided by following safeguarding measures such as perceived cost and self-efficacy. Research conducted by authors in [22, 23] has identified the co-relationship of self-efficacy with knowledge. For example, a user is more confident to take action against Phishing attack if they have knowledge on how to avoid phishing threats [12]. TTAT Theory shows that self-efficacy is an important factor in determining malicious IT threat avoidance motivation phenomenon. In addition, author in [31] has identified that knowledge can be influenced by learning procedural and conceptual knowledge in technical context. Consistent with this relationship authors in [12] derived a theoretical model based on TTAT model. The derived theoretical model describes how an individual IT users' threat avoidance behaviour is determined by avoidance motivation and affected by self-efficacy. The theoretical model investigated the effect of conceptual and procedural knowledge on user's self-efficacy to combat against phishing threats. Their study clearly showed that the combination of both conceptual and procedural knowledge positively affects user's confidence by enhancing their phishing threat avoidance behaviour. However, one can argue that knowledge cannot be limited within procedural and conceptual knowledge and there are other knowledge elements that help raise the user's self-efficacy. Expanding user's self-efficacy beyond conceptual and procedural knowledge will help them to protect themselves from any potential phishing attacks.

The main idea of 'know how to do it' knowledge is simply procedural knowledge [32]. Conceptual knowledge, on the other hand, is simply consists of ideas that give some power to think ('know that') in technological context [32]. Problem solving, process, strategic thinking are the ideas behind the procedural knowledge [31]. Whereas, conceptual knowledge is concerned with relationships among 'items' of knowledge [31]. The inter-relationship of conceptual and procedural knowledge is the idea to 'know-how-to-it-by-knowing-that'. However, it is not sufficient to know 'how' and 'that'. In order to know how, you must have the idea to know why to do that [25]. Therefore, these two (conceptual and procedural knowledge) knowledge learning attributes are not sufficient to enhance users' knowledge without knowing 'why' to apply these knowledge. Structural knowledge is nothing but the idea of 'know why'. Hence, structural knowledge is applicable in creation of plans and strategies, setting conditions for different procedures, and what to do when failure arises or when a part of information is missing [1]. Thus, the structural knowledge enhances users' self-efficacy. According to social cognitive theory, self-efficacy comes from four main sources: mastery experiences, observational learning, social persuasion and emotional arousal [15]. In addition, judgments of self-efficacy are the outcome of past experience, vicarious experience (i.e. the experience modelled by others), social persuasion/influence after training and feedback, and also from the physical and emotional state of individuals [2]. Therefore, both observational learning and past experience knowledge have a strong impact to increase individuals' self-efficacy.

## 3   Theoretical Model development

The literature review in the previous section has identified that knowledge learning attributes have influence on IT user's self-efficacy. The proposed theoretical model shown in Figure 1 is derived from the Arachchilage and Love's [12] theoretical model. The proposed model incorporates various knowledge elements to enhance users' self efficacy and describes how users' self efficacy influences users' IT threat avoidance motivation and IT threat avoidance behaviour. Self-efficacy is influenced by procedural knowledge, conceptual knowledge and their interaction along with structural knowledge, heuristic knowledge and observational knowledge. Lastly, the research reported in this paper attempted to examine whether all these knowledge learning attributes (structural knowledge, heuristic knowledge and observational knowledge) affects on IT users' self-efficacy to combat against phishing attacks.

The Hypotheses (H) are described as follows:
**H1**: Avoidance behavior is positively influenced by avoidance motivation.
**H2**: Avoidance motivation is positively influenced by self-efficacy.
**H3a**: Procedural knowledge positively influences self-efficacy.
**H3b**: Conceptual knowledge positively influences self-efficacy.
**H3c**: The combination of both procedural and conceptual knowledge positively influences self-efficacy.
**H3**: The combination of H3a, H3b and H3c along with structural knowledge positively influences self-efficacy.
**H4a**: Heuristic knowledge positively influences self-efficacy.
**H4b**: Observational knowledge positively influences self-efficacy.
**H4c**: The combination of heuristic knowledge and observational knowledge positively influences self-efficacy.

In the proposed theoretical model, knowledge elements such as observational knowledge, heuristic knowledge and structural knowledge have been included in the TTAT based theoretical model developed in [12]. Additional knowledge elements, as shown as dashed line, have been included to evaluate their effect on users' self- efficacy and hence users' ability to take actions against Phishing attacks after gaining these knowledge elements. The proposed model intended to determine the effect of knowledge elements and interaction between knowledge elements in enhancing users' self-efficacy. It also helps determine how these knowledge elements enhance the user's self-efficacy and motivate themselves to avoid phishing threat through their motivation. The proposed knowledge elements contribute to the users' learning process and build their confidence level. Ths earning phishing URLs and websites through observation, experience, and structural manner will improve IT users' self-efficacy.

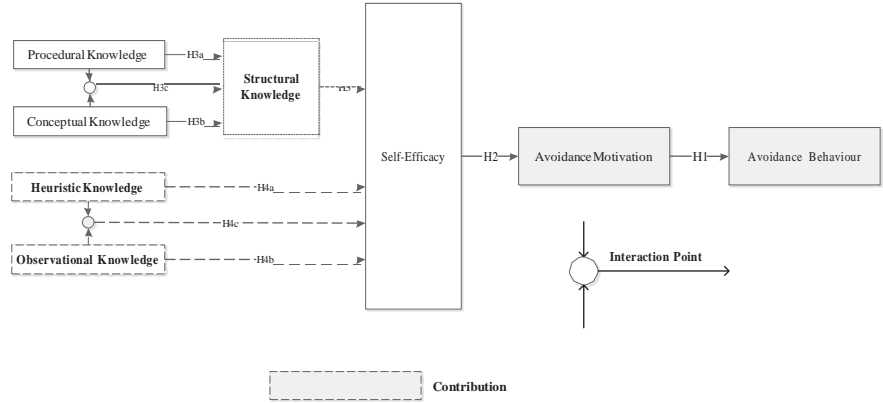

**Fig. 1.** Theoretical Framework.

### 3.1 Observational Knowledge

Knowledge is gained through a variety of ways such as by thinking, observing, experimenting, from science, theory and from experience [31]. Observational knowledge is the knowledge learned while observing a particular work or task being performed [21]. For example, when a new employee joins an organisation, the employee is assigned to a supervisor (i.e. an experienced employee), who works as a mentor. The new employee is asked to observe work being performed by others to gain knowledge (i.e. learning knowledge by observing peers' work). According to Bandura's social cognitive theory, people learn by observing others behavior without undergoing the trial and error process of operating it [16]. In addition to this, the social cognitive theory also explains that if individual's observed behavior produces expected result then the individual is more motivated to adopt the behavior and repeat it in the future [16]. Bandura's self-efficacy theory also suggests that one's ability to perform a desired outcome influences both thought and action process [16]. In general, the higher self-efficacy will act as a positive construct for taking action, whereas lower self-efficacy beliefs can have negative impact on the decision to execute a particular course of action [4]. With regard to technology in teaching and learning, user's self-efficacy is also influenced by thoughts and actions [4]. Observing a particular work is nothing but an action by individual. It is expected that high self-efficacy beliefs will function as a positive support to observe a particular course of action. Thus, the higher observational knowledge about phishing URLs/Emails will increase one's self-efficacy by increasing their phishing threat avoidance behavior. For example, one can observe the action on how to recognize phishing URLs/Emails and gain the observational knowledge about phishing websites and Emails. This knowledge will create phishing threat avoidance behavior and ultimately enhance their self-efficacy to protect themselves against phishing attacks. Therefore, the ob-servational knowledge contributes to the enhancement of observer's self-efficacy positively. Therefore, the hypothesis for this attribute is defined as follows:

**H4b**: Observational knowledge positively influences self-efficacy.

### 3.2   Heuristic Knowledge

Heuristic knowledge is the knowledge gained from past experience and is derived from previous experiences with similar problems [26]. These strategies rely on using readily accessible, though loosely applicable, information to control problem solving in human beings, machines, and abstract issues. For example, if a person is cheated while doing online transaction then if the same person needs to do another online transaction after some days, that individual will use the past experience (i.e. previous online transaction experience) to avoid being cheated again. According to social cognitive theory, people learn through their vicarious experience and this knowledge enables people to store the information required to guide their future behavior [16]. Self-efficacy beliefs not only influence decisions and behavior, but also these self-efficacy beliefs are influenced by other characteristics and prior experience within a particular domain [4]. Bandura in [17] described that past experiences have strong influence on self-efficacy beliefs and thus have a strong influence on behavior. Thus, the heuristic knowledge about phishing URLs/Emails will increase one's self-efficacy by using their phishing threat avoidance behaviour. Hence, heuristic knowledge is another source of an individual's self-efficacy. Therefore, the hypothesis for this attribute is defined as follows:

**H4c**: Heuristic knowledge positively influences self-efficacy.

### 3.3   Structural Knowledge

Self-efficacy is influenced by procedural and conceptual knowledge which is basically the idea to know "how" and know "that". However, it is not sufficient to know "that". In order to know how, you must have the idea to know 'why' [25]. Structural knowledge gives the idea to know why; it describes the rule sets, concept relationships, concept-to-object relationships [25]. For example, mango is a kind of fruit, fruit is a kind of crop and crop information is part of agricultural knowledge. Structural knowledge is basic to problem-solving and is the knowledge of how the ideas within a domain are integrated and interrelated. Procedural knowledge involves the idea how to do something. For example, by knowing how to differentiate between fake and real URLs/Emails (i.e., by knowing how the URLs/Emails are constructed) IT user can develop their procedural knowledge. Conceptual knowledge conveys the idea of understanding concepts and recognizing their application. For instance, IT user can learn the concept of features that fake and real URLs/Emails contain. Whereas, structural knowledge shows the idea of why to apply these concepts fake and real URLs/Emails of learning and how to recognize them (i.e., how-to-do-it-why-to-do-it-when). In general, Structural knowledge mediates the translation of conceptual knowledge into procedural knowledge and provides the application of procedural knowledge [25]. Therefore, having procedural, conceptual and structural knowledge will give the better idea to differentiate legitimate and illegitimate URLs/Emails. Thus the structural knowledge could be a factor in self-efficacy and this research intends to determine its' influence on users self-efficacy. The hypothesis for this attribute is defined as follows:

**H3**: The combination of H3a, H3b and H3c along with structural knowledge positively influences self-efficacy.

## 4 Integrating Theoretical Model into an Anti-Phishing Gaming Tool

The main focus of this research is to investigate how one can better educate people through various kind of knowledge representation (i.e., observational knowledge, heuristic knowledge and structural knowledge) into a gaming tool. The game design enhances users' self-efficacy to protect themselves from phishing attacks. The aim is also to demonstrate that while developing an anti-phishing learning tool, these knowledge factors can be taken into consideration for better education. The outcome will show that the knowledge attributes which enhances IT user's self-efficacy can be also be integrated into the anti-phishing learning tool. To explore this mechanism, a gaming prototype will be developed. A screen shot of the video game is shown in Figure **??**.

### 4.1 Game Story and Design

The game story is based on the game scenario of a balloon shooter. The main character of this game is Tom (i.e., the user or game player) who needs to shoot the balloons to play the game. Balloons are randomly generated in the game. Each balloon is associated with URLs/Emails which appear as a dialog box. Some balloons are associated with legitimate URLs/Emails and some of them are associated with fake URLs/Emails. To score well, the player (i.e., Tom) should be careful while shooting the balloons and Tom should shoot the balloons with real URLs/Emails. When Tom will point the gun towards the balloon, URL/Email associated with it will be displayed for the recognition of fake and legitimate ones. If the correct balloon (i.e., legitimate URLs/Emails) is shot by Tom then he will be awarded positive score and if not then he will be provided with "Facts-and-advice". Facts-and-advice is the opportunity for the game player to know about his/her mistake while recognising the URLs/Emails as summary and caution. It will help the player not to repeat the same mistake again. The other character of this proposed game is Jerry who plays the role of a guide and will provide the help to identify the URLs/Emails when asked by Tom (i.e., when Tom click the help button). Demonstration video (i.e., animated video) will be played before starting of the game to give a clear idea of how to play it. The proposed game design will have three different stages such as easy, medium and advance stage. The complexity will grow as the players are promoted from one stage to another stage. The proposed game design prototype determines how to identify legitimate URLs/Emails through the knowledge attribute (i.e., the knowledge gained by observation, experience and by knowing the structure of URLs/Emails).

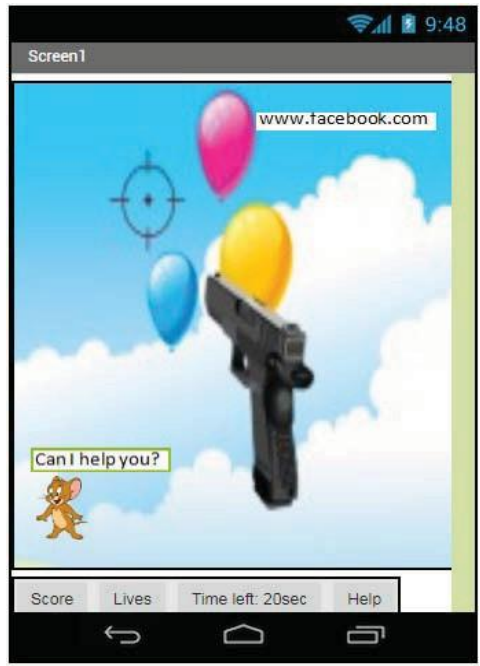

**Fig. 2.** Screen shot of the video game.

**Scoring Mechanism** For easy stage, a total of 10 balloons will be shown out of which there will be 7 and 3 balloons showing real and fake URLs/Emails respectively. To complete this stage the players will be given a total of 150 seconds (i.e., 15 second for each balloon). The number of lives given to a player will be 3. Each correctly shot balloon will add 5 points to the score and each wrong shot balloon will deduct 1 point from the score.

For the medium stage, a total of 15 balloons will be shown out of which there will be 8 balloons showing real URLs/Emails and 7 balloons showing fake URLs/Emails. Out of these 7 fake URLs/Emails displaying balloons, 3 balloons will display the fake URLS/Emails from the easy stage. To complete this stage the players will be given a total of 225 seconds (i.e., 15 seconds for each balloon). The number of lives given to a player will be 3. Each correctly shot balloon will add 5 points to the score and each wrong shot balloon will deduct 1 point from the score.

For the advance stage, a total of 20 balloons will be shown out of which there will be 10 balloons showing real URLs/Emails and 10balloons showing fake URLs/Emails. Out of these 10 fake URLs/Emails displaying balloons, 4 balloons will display the fake URLS/Emails from the medium stage. To complete this stage the players will be given a total of 300 seconds (i.e., 15 seconds for each balloon). The number of lives given to a player will be 3. Each correctly shot balloon will add 5 points to the score and each wrong shot balloon will deduct 1 point from the score.

**Game Design Scenario for Determining Observational Knowledge** To incorporate the observational knowledge of the user into this gamified approach, Tom (i.e., game player) will be asked to play the game. Firstly, the animated demo video of the game will be played in front Tom. The idea of playing this video is to learn how to differentiate between legitimate and illegitimate URLs/Emails by observing the demo video. Tom (game player) is not allowed to skip the demo video as it is important to learn how and what to observe. Each element will be able to be measured once game is developed. At that time, in order to gain observational knowledge, Tom will observe how the video shows tricks to identify real and fake URLs/Emails. After observation session is finished, Tom will be provided with a set of questions to check whether he has gained any knowledge about identifying URLs/Emails while observing demonstration of the game. Further, Tom will be asked to play the actual game after finishing a short training. After finishing the game Tom will be again provided by a questionnaire that will contains all the URLs/Emails. The idea behind this is to know how Tom uses his observational knowledge (i.e., why and how he marked the given URLs/Emails as fake or true) to play the game. This process will incorporate users' observational knowledge in to a game based learning tool. The score will be calculated through Tom's performance and the effect of observational knowledge on Tom will be checked and evaluated. This mechanism will determine whether or not observational knowledge influences users' self-efficacy to thwart against phishing attacks.

**Game Design Scenario for Determining Heuristic Knowledge** As mentioned before, the game will contain three different stages of difficulty such as easy, medium and advanced. Firstly, the demo video will be played and this will give an opportunity to Tom to gain some experience about how to recognize and differentiate real and fake URLs/Emails. After this the game player will play the easy stage first and the score will be recorded. After playing the easy stage, the same player will be asked to come and play the medium stage. The medium stage will contain some balloons (i.e., maximum 3 balloons) with same URLs/Emails as the easy stage. The result of the medium stage will be recorded. The third stage, which is advanced stage, will contain same balloons (i.e., four balloons) from both easy and medium stage. The result of the advanced stage will also be recorded. After Tom finishes playing all the stages of the game, the game player will be asked how he uses his experience of demo play while playing the game by himself. The recorded results will be analysed to check the effect of past experience (heuristic) knowledge later in the evaluation stage. This anal-ysis will be carried by comparing Tom's response to the balloons with same URLs/Emails for all stages of the game. This will show his experience gained from the first stage and its application in the next stage. The analysis would determine the performance of Tom while playing all stages of game with the same URLs/Emails. This game design scenario will enable us to check the effect of past experience knowledge.

**Game Design Scenario for Determining Structural Knowledge** To explore structural knowledge through game-based learning tool,firstly the demo video will be played in front of Tom. The demonstration video will show the concept behind fake and real URLs/Emails, procedure to identify them and basic problem solving knowledge to correctly recognize them (i.e., structural knowledge). In general this video will describe the structure of legitimate and illegitimate URLs/Emails along with the concept and procedure. In addition, this process teaches Tom how to identify the Phishing URLs and phishing emails through their structure (i.e., how they are constructed) and their particular features for easy recognition. After gaining this information, Tom will play the game and the outcome will be stored. After playing game, Tom will be asked to clarify his answer this means he will say how he recognize all the illegitimate and legitimate URLs/Emails by using his conceptual, procedural and structural knowledge. The outcome will be checked against his structural knowledge gained from demo video through a set of questionnaire in the evaluation phase. The set of the questionnaire will ask the game player how they used the structural knowledge to differentiate fake and legitimate URLs/Emails.

**Fact and Advice** After finishing playing a game, a "Fact and Advice" option will be made available to the player. The fact option will show the summary of the game that the player made mistakes. In addition to this, it will also tell the player what the mistake is and why it is a mistake (wrong decision). The advice option is basically the caution to the player and will show a set of guidelines on how to avoid the mistakes and try not doing the same mistake again in future. The provision of fact and advice to players can be considered as providing feedback to players. Players will learn new strategies for identifying phishing threats through the feedback they receive. When constructive feedback is provided to the players, it can affirm to players that they know what they are doing and that would eventually enhance their self-efficacy in phishing threat avoidance. The knowledge representation of this feature is to guide the game player on how to get observational, heuristic and structural knowledge and store them in their memory and apply them to protect themselves from phishing attacks.

## 5   Conclusion and Future Work

As we have become more and more dependent on Internet and its enabled devices for relaying our everyday life, we are becoming highly vulnerable to cyber-attacks and one of the dangerous attacks is known as Phishing. Many anti-phishing solutions have been proposed and developed to avoid and regulate phishing attacks and also to provide protection to users against such attacks. There is also a cat and mouse game between anti-phishing tool developers and cyber criminals (i.e. phishing attackers). Since we humans are the end user of the computer system and most vulnerable to phishing attacks, we can avoid being attacked by increasing our awareness of vulnerability and increasing our understanding and knowledge of phishing attack. Self-efficacy with its co-relation with knowledge is one of the most important determinants of individual's motivation in phishing threat avoidance behaviour. The proposed research endeavours on the user's self-efficacy in order to enhance the individual's phishing threat avoidance behaviour through their motivation. Using social cognitive theory, we explored that various knowledge attributes such as observational knowledge, heuristic knowledge and structural knowledge contributes immensely towards the individual's self-efficacy. Thus, this research aims to incorporate user's confidence to enhance their phishing awareness through their phishing threat avoidance motivation and behaviour. The proposed research contributes in the reduction in number of phishing attacks and enhancement of user confidence through their increased awareness and phishing avoidance knowledge.

The future work for this proposed research will empirically investigate the proposed theoretical model. With the research findings, a prototype of the game will be designed and developed with the aforementioned scenarios. A series of questionnaires will be developed and proper environment will be created to test the prototype of the game. At the time of writing of this paper, the type, structure and number of questionnaires to evaluate the prototype is still being worked upon. Finally, the last part of this research is to test the proposed theoretical model through the participants' engagement with the developed game prototype.